\documentclass{appolb}
\usepackage{epsfig}

\begin{document}

\title{Clusters in Light Nuclei 
\thanks{Presented at the 
Zakopane Conference on Nuclear Physics "Extremes of the Nuclear Landscape"
XLV in the series of Zakopane Schools of Physics -
International Symposium  - Zakopane, Poland, August 30 - September 5, 2010}
}

\author{C.~Beck$^a$, P.~Papka$^{a, g}$, A.~S\`anchez i Zafra$^a$, 
S.~Thummerer$^{a, b}$, F.~Azaiez$^{a, h}$, P.~Bednarczyk$^{a, i}$, 
S.~Courtin$^a$, D.~Curien$^a$, O.~Dorvaux$^a$, A.~Goasduff$^a$, 
D.~Lebhertz$^{a, j}$, A.~Nourreddine$^a$, M.~Rousseau$^a$, 
M.-D.~Salsac$^{a, k}$,
W.~von Oertzen$^b$, B.~Gebauer$^b$, C.~Wheldon$^{b, l}$, Tz.~Kokalova$^b$,
G.~Efimov$^{b, m}$, V.~Zherebchevsky$^{b, n}$, Ch.~Schulz$^b$, H.G.~Bohlen$^b$,
D. Kamanin$^{b, m}$, G.~de Angelis$^c$, A.~Gadea$^c$, S.~Lenzi$^c$, 
D.R.~Napoli$^c$, S.~Szilner$^{c, o}$, M.~Milin${^p}$, 
W.N.~Catford$^d$, D.G.~Jenkins$^e$, G.~Royer$^f$
\address{
$^a$ IPHC, IN$_{2}$P$_{3}$-CNRS/Universit\'e de Strasbourg, Strasbourg, France\\
$^b$ Helmholtz-Zentrum Berlin and Freie Universit\"at Berlin, Berlin, Germany\\
$^c$ INFN and Dipartimento di Fisica, Padova, Italy\\
$^d$ School of Physics and Chemistry, University of Surrey, Guildford, UK\\ 
$^e$ Department of Physics, University of York, York, UK\\ 
$^f$ Subatech, IN$_{2}$P$_{3}$-CNRS/Universit\'e de Nantes-EMN, Nantes, France\\
$^g$ Department of Physics, University of Stellenbosch, Stellenbosch, South Africa\\
$^h$ IPN Orsay, IN$_{2}$P$_{3}$-CNRS, Orsay, France\\
$^i$ The Institute of Nuclear Physics, Krak\'ow, Poland\\
$^j$ GANIL, IN$_{2}$P$_{3}$-CNRS/CEA, Caen, France\\
$^k$ CEA-Saclay, Irfu, Gif sur Yvette, France\\
$^l$ School of Physics and Astronomy, University of Birmingham, UK\\
$^m$ Joint Institut Nuclear Reactions, Dubna, Russia\\
$^n$ Saint Petersburg State University, Saint Petersburg, Russia\\ 
$^o$ Ruder Boskovic Institute, Zagreb, Croatia\\
$^p$ Faculty of Science, University of Zagreb, Zagreb, Croatia\\
         }
}

\maketitle

\newpage

\begin{abstract}

A great deal of research work has been undertaken in the $\alpha$-clustering 
study since the pioneering discovery, half a century ago, of 
$^{12}$C+$^{12}$C molecular resonances. Our knowledge of the field of 
the physics of nuclear molecules has increased considerably and nuclear 
clustering remains one of the most fruitful domains of nuclear physics, 
facing some of the greatest challenges and opportunities in the years ahead. 
In this work, the occurence of ``exotic'' shapes in light N=Z $\alpha$-like 
nuclei is investigated. Various approaches of superdeformed and hyperdeformed 
bands associated with quasimolecular resonant structures are presented. 
Results on clustering aspects are also discussed for light neutron-rich 
Oxygen isotopes.

\end{abstract}


\section{Introduction}

\begin{figure}[htb]
\begin{center}
\epsfxsize=120mm
\epsfbox{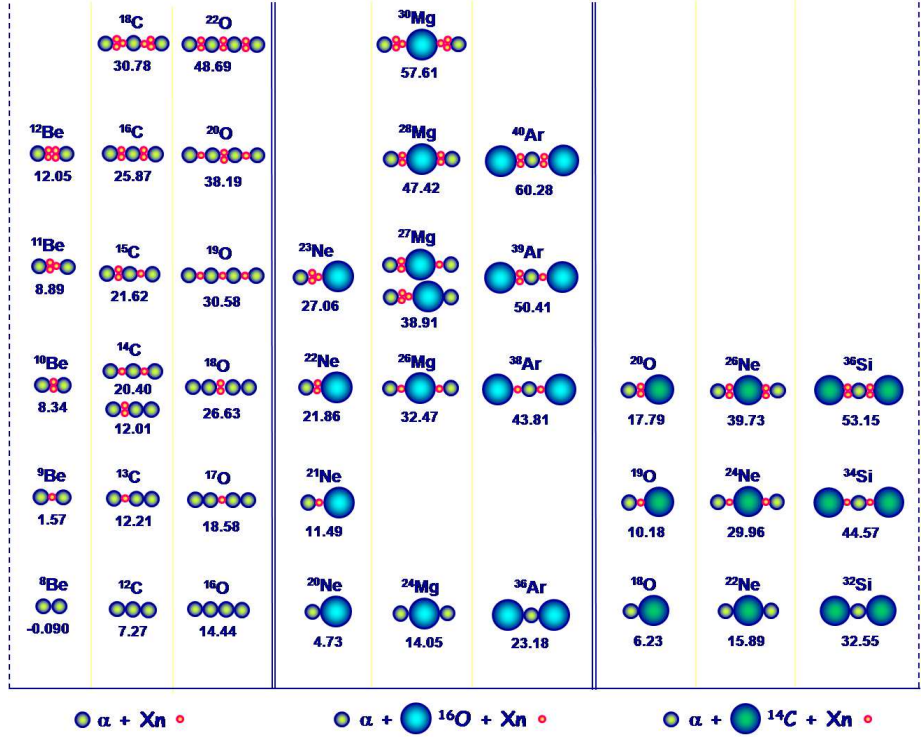}
\caption{(Color online) Schematic illustration of the structures of molecular
shape isomers in light neutron-rich isotopes of nuclei consisting
of $\alpha$-particles, $^{16}$O- and $^{14}$C-clusters plus some
covalently bound neutrons (Xn means X neutrons). The so called "Extended 
Ikeda-Diagram" \cite{Oertzen01} with $\alpha$-particles (left panel) and 
$^{16}$O-cores (middle panel) can be generalized to $^{14}$C-clusters cores 
(right panel). Threshold energies (in MeV) are given for the relevant 
decompositions.}
\end{center}
\end{figure}

The observation of resonant structures in the excitation functions for various 
combinations of light $\alpha$-cluster (N = Z) nuclei in the energy regime 
from the Coulomb barrier up to regions with excitation energies of E$_{x}$ = 20-50~MeV 
remains a subject of contemporary debate~\cite{Greiner95,Beck94}. These 
resonances have been interpreted in terms of nuclear molecules~\cite{Greiner95}. 
The question of how quasimolecular resonances may reflect contiuous transitions
from scattering states in the ion-ion potential to true cluster states in the 
compound systems is still unresolved \cite{Greiner95,Beck94}. In many cases, these 
resonant structures have been associated with strongly-deformed shapes and with 
alpha-clustering phenomena \cite{Freer07,Horiuchi10,Beck99}, predicted from the 
Nilsson-Strutinsky approach, the cranked $\alpha$-cluster model~\cite{Freer07}, or 
other mean-field calculations~\cite{Horiuchi10,Gupta10}. In light $\alpha$-like 
nuclei clustering is observed as a general phenomenon at high excitation energy 
close to the $\alpha$-decay thresholds \cite{Freer07,Oertzen06}. This exotic 
behavior has been perfectly well illustrated by the famous "Ikeda"-diagram for N=Z 
nuclei in 1968 \cite{Ikeda} that has been recently extended by von Oertzen 
\cite{Oertzen01} for neutron-rich nuclei, as shown in the left panel of Fig.1.
Clustering is a general phenomenon not only observed in light
neutron-rich nuclei \cite{Kanada10} but also in halo nuclei such as $^{11}$Li 
\cite{Ikeda10} or in very heavy systems where giant molecules can exist
\cite{Zagrebaev10}.

\section{Alpha clustering, deformations and alpha condensates}

The relationship between superdeformation (SD), nuclear molecules and alpha 
clustering \cite{Horiuchi10,Beck99,Beck04a,Beck04b,Cseh09} is of 
particular interest, since nuclear shapes with major-to-minor axis ratios of 
2:1 have the typical ellipsoidal elongation (with quadrupole deformation 
parameter $\beta_2$ $\approx$ 0.6) for light nuclei. Furthermore, the structure 
of possible octupole-unstable 3:1 nuclear shapes (with $\beta_2$ $\approx$ 
1.0) - hyperdeformation (HD) - for actinide nuclei has also been widely 
discussed \cite{Cseh09} in terms of clustering phenomena. Typical examples of 
a possible link between quasimolecular bands and extremely deformed (SD/HD) 
shapes have been widely discussed in the literature for light N = Z nuclei, 
for A$_{\small CN}$ = 20-60, such as $^{28}$Si 
\cite{Taniguchi09}, $^{32}$S 
\cite{Horiuchi10,Kimura04,Lonnroth10,Chandana10}, 
$^{36}$Ar \cite{Cseh09,Beck08a,Svensson00,Sciani09,Beck09}, $^{40}$Ca 
\cite{Ideguchi01,Rousseau02,Torilov04,Taniguchi07,Norrby10}, $^{44}$Ti 
\cite{Horiuchi10,Leary00,Fukada09}, $^{48}$Cr \cite{Salsac08} and $^{56}$Ni 
\cite{Beck99,Nouicer99,Rudolph99,Beck01,Bhattacharya02}.

\begin{figure}[htb]
\begin{center}
\epsfxsize=120mm
\epsfbox{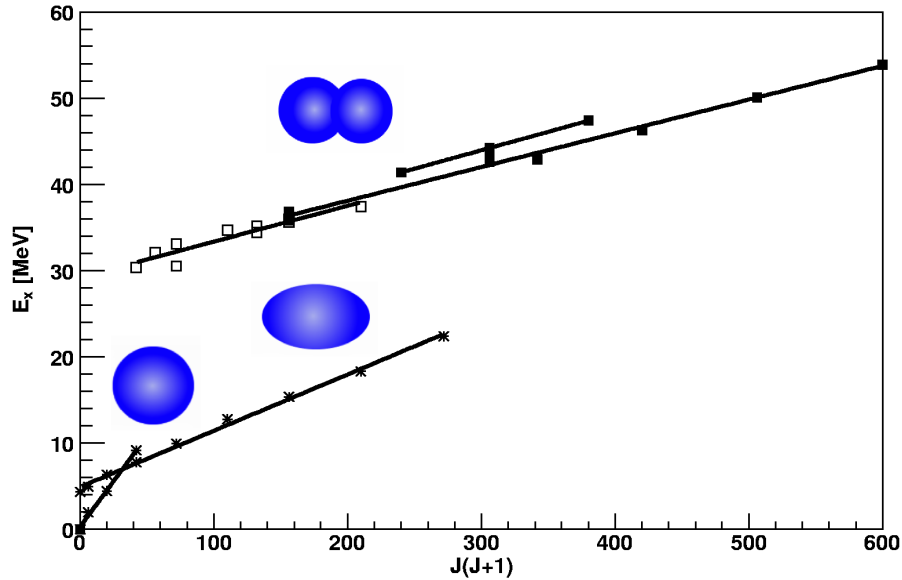}
\caption{Rotational bands and deformed shapes in $^{36}$Ar. Excitation energies 
	of the g.s. (spherical shape) and SD bands~\cite{Svensson00}
	(ellipsoidal shape), respectively, and the energies of HD band from 
	the quasimolecular resonances observed in the $^{12}$C+$^{24}$Mg 
	(open rectangles) \cite{Sciani09,Cindro79,Mermaz84,Pocanic85} and 
	$^{16}$O+$^{20}$Ne (full rectangles) \cite{Shimizu82,Gai84} reactions 
	(dinuclear shape) are plotted as a function of J(J+1). This figure
	has been adapted from Refs.~\cite{Beck08a,Sciani09}.}
\end{center}
\end{figure}

In fact, highly deformed shapes and SD rotational bands have been 
recently discovered in several such N = Z nuclei, such as $^{36}$Ar
and $^{40}$Ca by using $\gamma$-ray spectroscopy techniques 
\cite{Svensson00,Ideguchi01}. In particular, the extremely deformed rotational
bands in $^{36}$Ar \cite{Svensson00} (shown as crosses in Fig.~2) are observed 
as quasimolecular bands in both $^{12}$C+$^{24}$Mg 
\cite{Sciani09,Cindro79,Mermaz84,Pocanic85} (shown as open triangles in Fig.~2)
and $^{16}$O+$^{20}$Ne reactions \cite{Shimizu82,Gai84} (shown as full
rectangles), and their related ternary clusterizations are also predicted 
theoretically but were not found experimentally in $^{36}$Ar so far 
\cite{Beck09,Beck04c,Beck08b}. Similar negative results were previoulsy reported
for $^{48}$Cr \cite{Murphy96}.
On the other hand, ternary fission related to hyperdeformed 
shapes of $^{56}$Ni was identified from out-of-plane angular correlations 
measured in the $^{32}$S+$^{24}$Mg reaction with the Binary Reaction 
spectrometer (BRS) at the VIVITRON facility of the IPHC Strasbourg 
\cite{Oertzen08}. This finding \cite{Oertzen08} is not limited to light 
N = Z compound nuclei (note that same results were also obtained for $^{60}$Zn 
\cite{Oertzen08b,Zhere07}) but true ternary fission \cite{Zagrebaev10,Pyatkov10}
can also occur for very heavy \cite{Pyatkov10} and superheavy 
\cite{Zagrebaev10b} nuclei.

\begin{figure}[htb]
\begin{center}
\epsfxsize=60mm
\epsfbox{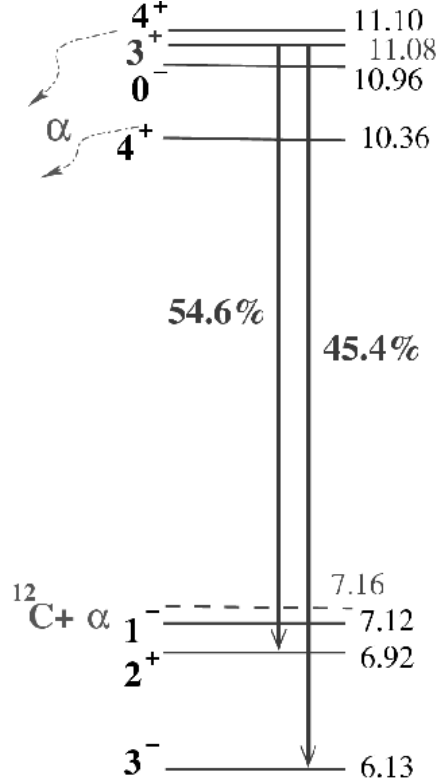}
\caption{New partial (high-energy) level scheme of $^{16}$O 
corresponding to $\gamma$-ray transitions observed in the
$^{12}$C($^{24}$Mg,$^{20}$Ne)$^{16}$O$^{*}$ $\alpha$-transfer 
reactions. This figure has been adapted from Ref.~\cite{Beck09}.}
\end{center}
\end{figure}

There is a renewed interest in the spectroscopy of the $^{16}$O nucleus at high 
excitation energy \cite{Beck09,Beck04c,Beck08b}. Exclusive data were collected with the 
inverse kinematics reaction $^{24}$Mg$+^{12}$C studied at E$_{lab}$($^{24}$Mg) 
= 130 MeV with the BRS in coincidence with the {\sc EUROBALL IV} installed at the 
{\sc VIVITRON} Tandem facility at Strasbourg \cite{Beck09,Beck04c,Beck08b}. From the 
$\alpha$-transfer reactions (both direct transfer and deep-inelastic orbiting 
collisions \cite{Sanders99}), new information has been deduced on branching ratios 
of the decay of the 3$^{+}$ state of $^{16}$O at 11.085~MeV $\pm$ 3 keV. The 
high-energy level scheme of $^{16}$O indicates in Fig.~3 that this state does not 
$\alpha$-decay because of non-natural parity (in contrast to the two neighbouring 
4$^{+}$ states at 10.36~MeV and 11.10~MeV, respectively) it decays also to the 
2$^{+}$ state at 6.92~MeV (54.6 $\pm$ 2 $\%$). By considering all the four 
possibilities of transitions types of the 3$^{+}$ state (i.e. E$_1$ and M$_2$ for 
the 3$^{+}$ $\rightarrow$ 3$^{-}$ transition and, M$_1$ and E$_2$ for the 3$^{+}$ 
$\rightarrow$ 2$^{+}$ transition), our calculations yield the conclusion that a value 
for the decay width $\Gamma_{\gamma}$ is fifty times lower than known previously, it 
means $\Gamma_{3^+}$ $<$ 0.23 eV. This result is important as it is the last known 
$\gamma$-decay level for the well studied $^{16}$O nucleus \cite{Beck09}. The highly 
collective state 3$^{-}$ has symmetry, which may not be realized by the 
$\alpha$-cluster structure. Hence, it has small overlap with the $\alpha$-cluster
dominated states and small alpha decay rate. In Section 3 we will discuss clustering 
effects in the other light neutron-rich oxygen isotopes: $^{17,18,19,20}$O.

In the framework of the study of Bose-Einstein Condensation (BEC) the
$\alpha$-particle state \cite{Tohsaki01,Oertzen10a} in light N=Z nuclei, 
an experimental signature of BEC in $^{16}$O, is at present of highest 
priority. An equivalent $\alpha$+''Hoyle" state \cite{Hoyle54} in $^{12}$C is 
predicted to be the 0$^{+}_{6}$ state of  $^{16}$O at about 15.1 MeV, which 
energy is just lying (i.e. $\approx$ 700 keV) above the 4$\alpha$-particle 
breakup threshold \cite{Funaki08}. However, any state in $^{16}$O equivalent 
to the so-called ''Hoyle" state \cite{Hoyle54} in $^{12}$C is most certainly 
going to decay by particle emission with very small, probably un-measurable, 
$\gamma$-decay branches. Very efficient particle-detection techniques will 
have to be used in the near future as such BEC states will be expected to 
decay by alpha emission to the ''Hoyle" state, and could be associated with 
resonances in $\alpha$-particle inelastic scattering on $^{12}$C leading to 
that state, or be observed in $\alpha$-particle transfer to the 
$^{8}$Be-$^{8}$Be final state. Another possibility might be to perform Coulomb 
excitation measurements with intense $^{16}$O beams at intermediate energies.

\section{Clustering in light neutron-rich nuclei}

As discussed previously clustering is a general phenomenon also observed for 
nuclei with extra neutrons in an extended "Ikeda"-diagram \cite{Ikeda} as 
proposed by von Oertzen \cite{Oertzen01} (see the left panel of Fig.~1). With 
additional neutrons specific molecular structures appear, with binding effects 
based on covalent molecular neutron orbitals. In these diagram $\alpha$-clusters 
and $^{16}$O-clusters (as shown by the middle panel of the diagram of Fig.~1) are 
the main ingredients. Actually, the $^{14}$C nucleus has equivalent properties 
as a cluster, as the $^{16}$O nucleus: i) closed neutron p-shells, ii) first 
excited states well above E$^{*}$ = 6 MeV, and iii) high binding energies for 
$\alpha$-particles.

\begin{figure}[htb]
\begin{center}
\epsfxsize=120mm
\epsfbox{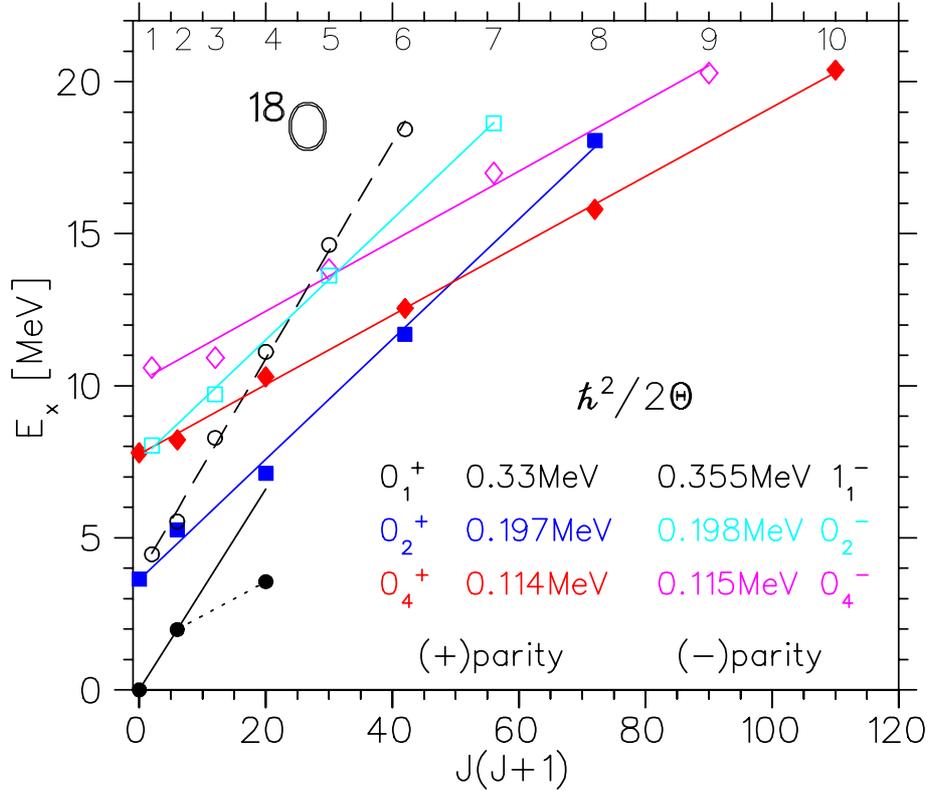}
\caption{(Color online) Overview of six rotational band structures observed in 
	$^{18}$O. Excitation energy systematics for the members of the rotational
	bands forming inversion doublets with K=0 are plotted as a function 
	of J(J+1). The indicated slope parameters contain information on
	the moments of inertia. Square symbols correspond to cluster bands,
	whereas diamonds symbols correspond to molecular bands. This figure
	is adapted from \cite{Oertzen10b}.}
\end{center}
\end{figure}

A general picture of clustering and molecular configurations in light nuclei 
can be drawn from a detailed investigation of the light oxygen isotopes with
A $\geq$ 17. Here we will only present recent results of the even-even 
oxygen isotopes: $^{18}$O \cite{Oertzen10b} (note that $\alpha$-cluster properties
of $^{18}$O are also discussed in detail in Ref.~\cite{Johnson09}) and $^{20}$O 
\cite{Oertzen09}. But very strinking cluster states have also been found in 
odd-even oxygen isotopes such as: $^{17}$O \cite{Milin09} and $^{19}$O \cite{Oertzen11}. 

Fig.~4 gives an overview of all bands in $^{18}$O as a plot of excitation energies
in dependence of J(J+1) together with their respective moment of inertia. In the
construction of the bands both the excitation energy systematics of the J(J+1)
dependence and the measured cross sections \cite{Oertzen10b} dependence on (2J+1)
were used. The experimental molecular bands are supported by either
generator-coordinate-method \cite{Descouvemont} or Antisymmetrized Molecular
Dynamics (AMD) calculations \cite{Furutachi08}. Slope parameters obtained in
a linear fit to the excitation energies data \cite{Oertzen10b} indicate the moment
of inertia of the rotational bands given in Fig.~4. The intrinsic structure
of the cluster bands is reflection asymmetric, the parity projection gives an 
energy splitting between the partner bands. 

For $^{20}$O \cite{Oertzen09}, we can compare the bands of Fig.~5 with those of $^{18}$O
(displayed in Fig.~4). The first doublet (K=0$^{\pm}_{2}$) has a moment of
inertia which is slightly larger (smaller slope parameter), consistent
with the interpretation as ($^{14}$C-$^{6}$He or $^{16}$C-$^{4}$He)  
molecular structures (they start well below the thresholds of 16.8 MeV and 
12.32 MeV, respectively). The second band, for which the negative parity 
partner is yet to be determined, has a slope parameter slightly smaller
as compared to $^{18}$O. This is consistent with  the study of bands in 
$^{20}$O by Furutachi et al. \cite{Furutachi08} that clearly establishes 
parity inversion doublets predicted by AMD calculations for the 
$^{14}$C-$^6$He cluster and $^{14}$C-2n-$\alpha$ molecular structures.

\begin{figure}[htb]
\begin{center}
\epsfxsize=120mm
\epsfbox{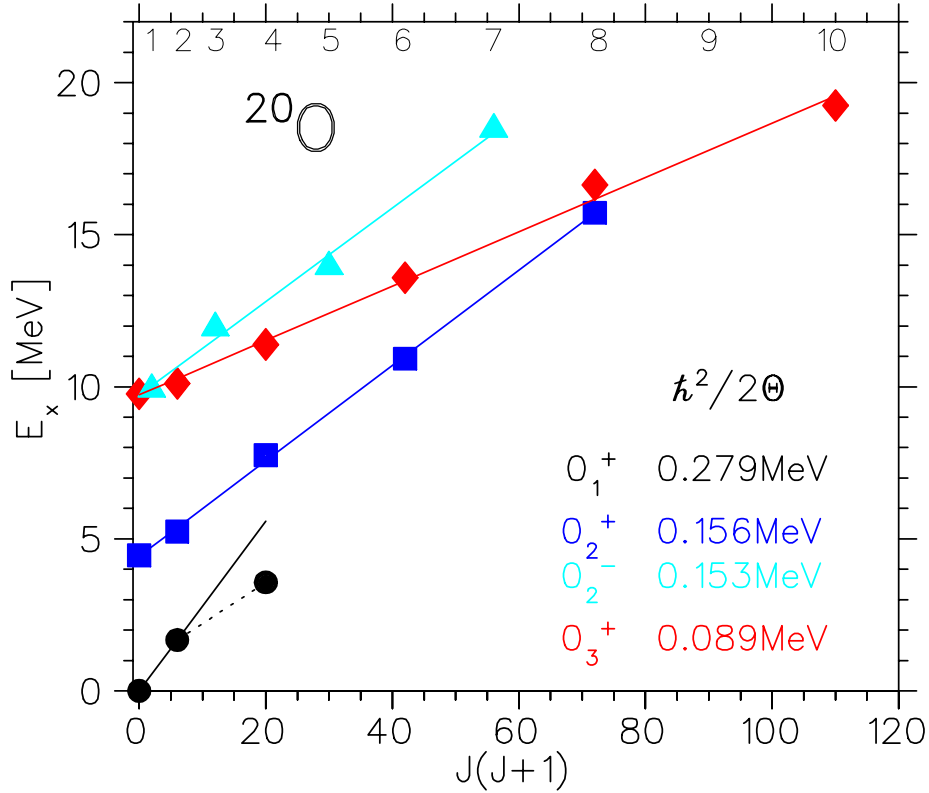}
\caption{(Color online) Overview of four rotational band structures observed in 
	$^{20}$O.
	Excitation energy systematics for the members of the rotational
	bands forming inversion doublets with K=0 are plotted as a function 
	of J(J+1). The indicated slope parameters contain information on
	the moments of inertia. Square and triangle symbols correspond to 
	cluster bands, whereas diamonds symbols correspond to molecular 
	bands.This figure is adapted from \cite{Oertzen09}}
\end{center}
\end{figure}

The corresponding momenta of inertia values given in Fig.~4 and Fig.~5 are 
highly suggesting large deformations of the clusters structures. We may also 
conclude that the strongly bound $^{14}$C nucleus has equivalent properties 
as an $^{16}$O cluster, as mentioned previously. Therefore, the Ikeda-diagram 
\cite{Ikeda} and the "extended Ikeda-diagram" consisting of $^{16}$O clusters 
cores with covalently bound neutrons \cite{Oertzen01} must be revised to 
include also the $^{14}$C clusters cores as illustrated by Fig.~1. 

\section{Summary and conclusions }

The connection of $\alpha$-clustering, quasimolecular resonances, orbiting 
phenomena and extreme deformations (SD, HD, ...) has been discussed in this 
work by using $\gamma$-ray spectroscopy of coincident binary fragments from 
either inelastic excitations and direct transfers (with small energy damping 
and spin transfer) or from orbiting (fully damped) processes \cite{Sanders99}.
From a careful analysis of the $^{16}$O+$^{20}$Ne 
$\alpha$-transfer exit-channel (strongly populated by orbiting) new information 
has been deduced on branching ratios of the decay of the 3$^{+}$ state of $^{16}$O 
at 11.089~MeV. This result is encouraging for a complete $\gamma$-ray spectroscopy 
of the $^{16}$O nucleus at high excitation energy. Of particular interest is the 
quest for the corresponding 4$\alpha$ states near the $^{8}$Be+$^{8}$Be and 
$^{12}$C+$\alpha$ decay thresholds. The search for extremely 
elongated configurations (HD) in rapidly rotating medium-mass nuclei, which has 
been pursued by $\gamma$-ray spectroscopy measurements, will have to be 
performed in conjunction with charged-particle techniques in the near future 
(see \cite{Oertzen08,Wheldon08}). In addition, we have presented new results
of neutron-rich oxygen isotopes displaying very well defined molecular bands
in agreement with AMD predictions. Consequently, the extended Ikeda diagram
has been revised for light neutron-rich nuclei to include the $^{14}$C 
cluster, similarly to the $^{16}$O cluster.\\

\section{Acknowledgments}

We thank the {\sc VIVITRON} staff and the EUROBALL group of 
Strasbourg for their excellent support. One
of us (C.B.) would like to acknowledge C. Bhattacharya, J. Cseh, 
Y. Kanada-En'yo, M. Norrby, and Y. Taniguchi for providing him with new 
results prior to their publication, as well as Christian Caron (Springer) for 
initiating the new series of Lecture Notes in Physics dedicated to "Clusters 
in Nuclei". 
This work was supported by the french IN2P3/CNRS, the german 
ministry of research (BMBF grant under contract Nr.06-OB-900), and the EC 
Euroviv contract HPRI-CT-1999-00078.

\end{document}